\documentclass[letterpaper]{article}
\pdfoutput=1 

\usepackage{jheppub} 

\usepackage[export]{adjustbox}
\usepackage[T1]{fontenc} 
\usepackage{graphicx} 
\usepackage{dcolumn} 
\usepackage{bm} 
\usepackage{amssymb} 
\usepackage{amsmath} 
\usepackage{amsfonts}
\usepackage{mathtools}
\usepackage{color}
\usepackage{comment}
\usepackage{rotating}
\usepackage{multirow}
\usepackage{ctable}
\usepackage{colortbl}
\usepackage{afterpage}
\usepackage{slashed}
\usepackage{array}

\definecolor{darkgreen}{RGB}{50,150,0}

\newcommand{\dd}{\mathrm{d}}
\newcommand{\pd}{\partial}
\newcommand{\ee}{\mathrm{e}}

\begin{document}
\title{A Tale of Two $U(1)$'s: Kinetic Mixing from Lattice WGC States}
\author{Georges Obied and Aditya Parikh}
\affiliation{Department of Physics, Harvard University, Cambridge, MA 02138, USA}

\abstract{We point out that the states required by the Lattice Weak Gravity Conjecture, along with certain genericity conditions, imply the existence of non-vanishing kinetic mixing between massless Abelian gauge groups in the low-energy effective theory. We carry out a phenomenological estimate using a string-inspired probability distribution for the masses of superextremal states and compare the results to expectations from string theory and field theory, estimating the magnitude of kinetic mixing in each case. In the string case, we compute the kinetic mixing in an ensemble of 1858 MSSM-like heterotic orbifolds as well as in Type II supergravity on a Calabi-Yau manifold. From the field theory perspective, we consider compactifications of a 5D gauge theory. Finally, we discuss potential loopholes that can evade the bounds set by our estimates.}

\maketitle

\section{Motivation and Introduction}
\label{sec:intro}
One generic feature of string compactifications is the presence of hidden sectors containing additional matter that is possibly charged under extensions of the Standard Model (SM) gauge group $G_{\rm SM}$. This expectation is borne out in our universe through the existence of a dark sector. In fact, cosmological observations have shown that about 95\% of the energy density in the universe resides within such a dark sector, with 25\% as dark matter and 70\% as dark energy~\cite{Aghanim:2018eyx}. Despite this abundance of dark matter, it has proven difficult to detect any non-gravitational interactions between our sector and the dark sector. It is a logical possibility that dark matter is not charged under any gauge group or that interactions between the two sectors are only mediated by operators suppressed by a large mass scale. In these cases, directly observing any hidden sector physics would require high energy experiments beyond the reach of current technology. However, if dark matter is charged under an additional $U(1)_X$, there is a dimension four kinetic mixing operator between $U(1)_X$ and the SM hypercharge $U(1)_Y$ that is not suppressed by a large mass scale:
\begin{align*}
    \mathcal{L} \supset -\frac{1}{4}F_{(Y)}^2 -\frac{1}{4}F_{(X)}^2
    + \frac{\chi}{2}F^{\mu\nu}_{(Y)}F_{(X)\mu\nu}
\end{align*}

\begin{figure}[tp]
    \centering
    \includegraphics[width=0.75\textwidth]{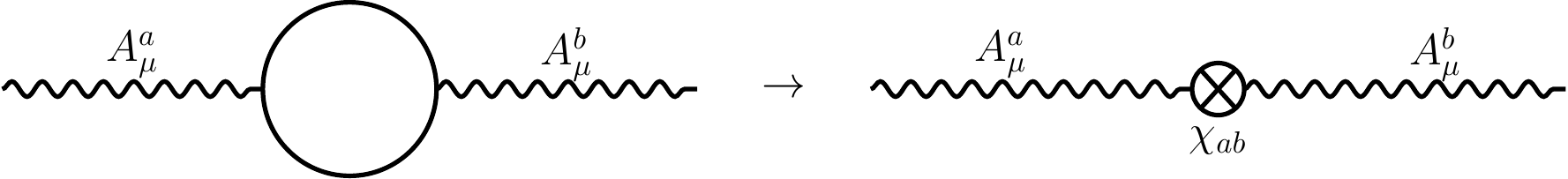}
    \caption{If bi-charged matter exists in the UV model, integrating it out can generate a 1-loop kinetic mixing in the IR via this diagram.}
    \label{fig:kin_mix_loop}
\end{figure}

This operator can be generated by integrating out massive particles in a UV theory that are charged under both Abelian groups~\cite{Holdom:1985ag} (see also~\cite{Fabbrichesi:2020wbt} for a review), as shown schematically in Figure~\ref{fig:kin_mix_loop}. Each scalar bi-charged state, with mass $m$ and charges $Q_{a}$ and $Q_{b}$ under the respective U(1)'s, contributes an amount $\chi$ to the kinetic mixing given by\footnote{The numerical prefactor changes if we integrate out fermions or chiral supermultiplets in a supersymmetric theory, but the dependence on charges and masses remains.}
\begin{equation}
    \frac{\chi}{g_{a}g_{b}} = -\frac{Q_{a}Q_{b}}{48\pi^{2}}\ln \frac{m^{2}}{\mu^{2}}
\label{eqn:kinmix_formula}
\end{equation}

The effect is similar to that of a threshold correction to the gauge coupling.
If the low-energy theory does not contain any more particles charged under $U(1)_X$, one is free to define a rotation in field space $A'_Y \equiv A_Y - \chi A_X$ along with a new gauge coupling $g'_Y\equiv g_Y/\sqrt{1-\chi^2}$ that removes all effects of the kinetic mixing operator. But in the presence of light charged hidden matter, the effect of a kinetic mixing operator is to give certain dark matter species a `minicharge' under the visible $U(1)_Y$ and in turn $U(1)_{\rm EM}$. This can be seen by a similar field redefinition as above which causes a dark matter particle of charge $g_X Q$ under $U(1)_X$ to acquire a charge $\chi Q g_X/g_Y$ under $U(1)_Y$. So far, we have assumed that the $U(1)_X$ gauge boson is also massless, which provides the freedom to rotate the mixing into either sector. An interesting alternative is to consider a massive dark photon. In this scenario, diagonalizing to the mass basis fixes the eigenstates and uniquely determines the coupling of the photon and dark photon to both the Standard Model and dark currents. The standard mechanism for generating a mass for the dark photon is by spontaneously breaking the symmetry with a Higgs or by providing a Stueckelberg mass, which is a special case of the Higgs mechanism where the Higgs has been decoupled. In either case, above the symmetry breaking scale, we have a massless Abelian field, so for the purposes of our analysis, we will restrict our attention to this case, with the expectation that our results will also apply to the massive case as well. The phenomenological implications of the massive dark photon have been extensively explored in recent years, so we will, of necessity, refer the readers to an excellent recent review article and the references therein~\cite{Fabbrichesi:2020wbt}.

Numerous current and near-future experimental efforts (see for example~\cite{Fabbrichesi:2020wbt} for a recent review) are directed at detecting minicharged particles. These have phenomenological signatures that vary with the mass of the dark matter particle and the strength of the kinetic mixing parameter $\chi$. For masses below $\sim 0.1\,\mathrm{MeV}$, the most constraining effect is due to stellar energy loss. This has been investigated using data from the Sun~\cite{Vinyoles:2015khy}, white dwarves, red giants and horizontal branch stars~\cite{Vogel:2013raa}. SN1987a also provides constraints extending to $\sim 1$ MeV~\cite{Chang:2018rso}. At higher masses, cosmological observations (for $\chi \lesssim 10^{-3}$) and laboratory searches (for larger $\chi$) are more relevant. The former rely on $N_{\rm eff}$ constraints from BBN and the CMB~\cite{Vogel:2013raa} as well as effects of minicharged particles on acoustic oscillations during recombination~\cite{Burrage:2009yz}.
The latter involve collider searches~\cite{Jaeckel:2012yz,Davidson:2000hf}, beam-dump experiments~\cite{Prinz:1998ua}, ortho-positronium decays~\cite{Badertscher:2006fm}, light shining through a wall experiments~\cite{Ehret:2010mh}
and neutrino experiments~\cite{Magill:2018tbb}. Future experiments will further explore this parameter space~\cite{Raubenheimer:2018mwt,Akesson:2640784,Akesson:2018vlm,Ball:2016zrp,Kelly:2018brz,Gninenko:2640930,Gninenko:2300189}.

Given the vast parameter space, it is important to provide theoretical input on the expected magnitude of the coefficient $\chi$. Since the value of $\chi$ in any effective field theory (EFT) is very model-dependent, one may hope that turning to a UV complete theory (with a more rigid framework) would provide a better picture of generic expectations. Indeed, several such investigations have been carried out in the literature giving predictions for the value of $\chi$ ranging over many orders of magnitude with some models having a seemingly vanishing mixing parameter.\footnote{Although the expectation is that mixing is present anyways from low-energy effects such as SUSY breaking, GUT symmetry breaking or Dine-Seiberg-Witten vacuum shifts.}

One of the pioneering studies is due to~\cite{Dienes:1996zr} where contributions to $\chi$ from massive string states were computed in semi-realistic orbifold compactifications of the heterotic string in the fermionic formulation. The surprising result was that massive string states do not contribute to mixing in the low-energy theory. Later works~\cite{Goodsell:2011wn} outlined the requirements for kinetic mixing to be present in heterotic orbifolds by extending early results about threshold corrections~\cite{Kaplunovsky:1987rp,Dixon:1990pc}. When present, the kinetic mixing effect is considerable with $\chi\gtrsim 10^{-4}$. In the context of Type II theories with D-branes, kinetic mixing has been found in Calabi-Yau orientifolds with D7-branes~\cite{Jockers:2004yj}, D6-branes~\cite{Abel:2008ai}, D5-branes~\cite{Grimm:2008dq} and in the LARGE volume scenario~\cite{Goodsell:2009xc}. Additionally, kinetic mixing can occur in non-SUSY set-ups with anti-branes in the extra dimensions~\cite{Abel:2003ue,Abel:2006qt} such as $\overline{\mathrm{D}3}$-branes in Klebanov-Strassler throats. Finally, kinetic mixing has also been shown to arise at strong coupling in F-theory constructions~\cite{DelZotto:2016fju}.

As such, different limits of the string landscape have been investigated but no overall picture is currently available. Here we revisit some of these arguments in light of the Sublattice Weak Gravity Conjecture (sLWGC) which is expected to hold in the entire landscape~\cite{Heidenreich:2015nta,Heidenreich:2016aqi}. As we review in the next section, the sLWGC requires massive states charged under all Abelian symmetries. When integrated out, these can generate the kinetic mixing operator in the low-energy EFT.
Of course, explicit constructions include these states but the highly symmetric set-ups required for computational control can possibly lead to large cancellations and thus biases estimates of $\chi$ towards lower values.
In this paper, we point out that the sLWGC, in addition to an assumption about the mass spectrum of relevant states, also leads to an estimate of $\chi$, extending the analysis of~\cite{Benakli:2020vng}. This approach is complementary to the available constructions in the sense that it is more general but inevitably less precise. The question of computing kinetic mixing effects is replaced by an estimate of the mass spectrum of states charged under the various $U(1)$ symmetries.

This paper is organized as follows. In Section~\ref{sec:swampland_review}, we begin with a review of the relevant Swampland conjectures. Next, we discuss a statistical approach to estimating the kinetic mixing arising from a generic spectrum of sLWGC states in Section~\ref{sec:sLWGC_stat}. We discuss specific constructions in Section~\ref{sec:constructions}. This includes a QFT construction in Section~\ref{sec:5D_example}, heterotic orbifold constructions in Section~\ref{sec:orbifolds}, and Type II on a Calabi-Yau in Section~\ref{sec:quintic}. Section~\ref{sec:loopholes} discusses various potential known loopholes. Finally, we offer concluding remarks in Section~\ref{sec:conclusions}.

\section{Swampland/Quantum Gravity Preliminaries}
\label{sec:swampland_review}
The Swampland conjectures are a set of conjectures which impose constraints on low-energy effective theories. These constraints arise from the fact that seemingly innocuous effective field theories in the IR run into pathologies when we try to UV complete them into a theory of quantum gravity. Theories satisfying these constraints are said to be in the Landscape, while theories violating these constraints live in the Swampland. Taking these conjectures as a theoretical input, we can make predictions for the size of certain parameters in our effective field theory, such as the kinetic mixing parameter $\chi$. These predictions allow us to bring quantum gravity into contact with our rich experimental program. In this section, we will review the relevant Swampland conjectures. See the reviews~\cite{Brennan:2017rbf,Palti:2019pca} for more details.

\subsection{Review of sLWGC}

The Weak Gravity Conjecture (WGC)~\cite{ArkaniHamed:2006dz} is one of the oldest conjectures in the Swampland program. The WGC states that if we have a U(1) gauge theory coupled to gravity, then there exists a (superextremal) particle whose charge to mass ratio is larger than that of a large, non-rotating, extremal black hole~\cite{ArkaniHamed:2006dz}.

The conjecture was originally proposed to avoid the existence of stable black holes. If there is a particle in the spectrum of states satisfying the WGC, then an extremal black hole can decay away from extremality by emitting a superextremal particle. Although appealing, a priori there isn't any fundamental obstruction in having a stable extremal black hole charged under a gauge symmetry in the same way that there is for a global symmetry.
On the other hand, we know the WGC holds in a large class of string theory constructions. Furthermore, simple toy examples such as Kaluza-Klein (KK) reductions also generate towers of states satisfying the WGC.

The authors of~\cite{Heidenreich:2015nta,Heidenreich:2016aqi} further refined this conjecture.
They generalized the Weak Gravity Conjecture to state that an Abelian $p$-form gauge field with gauge coupling $e_{p;d}$ in $d$ dimensions with varying dilaton couplings $\alpha$, necesitates the existence of a ($p-1$)-brane of tension $T_{p}$ and charge $q$ satisfying
\begin{equation}
    \Bigg[\frac{\alpha^{2}}{2} + \frac{p(d-p-2)}{d-2}\Bigg]T^{2}_{p} \leq e^{2}_{p;d}q^{2}M^{d-2}_{d}
\label{eqn:gen_WGC}
\end{equation}
They also proposed the Sublattice Weak Gravity Conjecture. The sLWGC states that if we have a charge lattice $\Gamma$ in a given theory, then there exists a sublattice $\Gamma_{\text{sup}}\subseteq\Gamma$ such that all $\vec{q}\in\Gamma_{\text{sup}}$ correspond to a superextremal state. The coarseness of the sublattice, defined as the smallest integer $N$ such that $N\vec{q}\in\Gamma_{\text{sup}}, \forall\vec{q}\in\Gamma$, is finite.
Therefore, $\Gamma_{\text{sup}}$ and $\Gamma$ have the same dimensionality. Crucially, in a U(1)$\times$U(1) gauge theory, the sLWGC implies the existence of a sublattice of superextremal states charged under both gauge groups. The existence of these states is an important starting point for generating kinetic mixing. A toy model illustrating this feature is the KK compactification of 6D pure gravity on $T^{2}$. This dimensional reduction generates two graviphotons, giving us a U(1)$\times$U(1) gauge theory in 4D. The masses of the KK modes are given by
\begin{equation}
    m^{2} = \frac{\phi^{ij}n_{i}n_{j}}{R^{2}} \quad \quad Q_{i} = n_{i}
\end{equation}
$\phi$ denotes the metric on moduli space and $R^{2}$ is the overall compactification scale. The extremality bound, which was also derived in~\cite{Heidenreich:2015nta}, is given by
\begin{equation}
    M^{2} \geq \frac{\phi^{ij}Q_{i}Q_{j}}{R^{2}}
\end{equation}
The KK modes saturate this bound. Furthermore, we have a state for each choice of $Q_{i}$. So, we immediately see that the sLWGC is satisfied with $\Gamma_{\text{sup}} = \Gamma$.\footnote{This was intended as an illustrative example. If we compactify on other manifolds, such as toroidal orbifolds, we are only guaranteed a sublattice of superextremal states.}
\subsection{Species Scale}
\label{sec:species_scale}
The QFT we analyze is an effective lower dimensional theory with gravity propagating in higher dimensions. This effective theory is valid up to a cutoff scale which is set by the species scale $\Lambda_{s}$. Above the species scale, gravity becomes strongly coupled in the higher dimensional theory. The species scale conjecture states that if we consider a $d$-dimensional gravitational theory having $N$ single particle states, with Planck mass $M_{p}^{(d)}$, then requiring gravity to remain weakly coupled necessitates a cutoff scale $\Lambda$ below the species scale given by
\begin{equation}
    \Lambda_{s} = \frac{M_{p}^{(d)}}{N^\frac{1}{d-2}}.
\end{equation}
This conjecture can be derived by counting KK modes, but is expected to hold more generally. For computing kinetic mixing, this implies that every light state we integrate out, lowers the cutoff of our theory. States with masses below $\Lambda_{s}$ contribute to kinetic mixing, while states with masses above are not part of our effective description and hence are excluded.

\section{Estimating $\chi$ from sLWGC states}
\label{sec:sLWGC_stat}
In this section, we carry out a phenomenological estimate of $\chi$ by assuming that the sLWGC holds in a high energy theory and integrating out string-scale states to obtain the low-energy kinetic mixing coefficient. This also allows us to investigate the dependence of the mixing on the mass distribution of superextremal particles on the charge lattice. To isolate the effect of these particles, we assume that the $U(1)$'s under consideration do not kinetically mix in the full theory. In addition, we assume that this mixing cannot be rotated away by redefining the $U(1)$ generators. This could be due to the existence of a light particle carrying either $U(1)$ charge for example, as mentioned in Section~\ref{sec:intro}, or other effects such as the possibility of measuring the coupling constant of a GUT. The $\chi$ estimates obtained here should be viewed as lower bounds since we are only considering superextremal states whereas any string construction typically includes a large number of subextremal states as well.

More precisely, we consider the charge lattice of a $U(1)_a \times U(1)_b$ theory with superextremal particles of mass:
\begin{equation}
    \frac{m}{M_{\text{Pl}}} = c \times q \qquad\quad q \equiv \sqrt{(g_a Q_a)^{2} + (g_b Q_b)^{2}}
\label{eqn:WGC_mass}
\end{equation}
at position $(Q_a, Q_b)$ on the lattice. The expression for $q$ in Equation~\ref{eqn:WGC_mass} is a good approximation when the  mixing is small. However when the mixing is $\mathcal{O}(1)$, the off-diagonal elements of the metric on field space contribute to this formula. We will ignore this subtlety for our estimate. The $c$'s are random coefficients drawn from a distribution that depends on the charge site and are strictly less than unity to ensure superextremality. In principle, the probability density function of $c$ is calculable from Landscape constructions but this is prohibitively difficult in practice except perhaps within the context of a limited class of models.\footnote{Assigning a probability distribution to $c$ is akin to assigning a probability measure on theory space and we are certainly not claiming that we have access to that measure. Instead one may think of this probability distribution as an area weight over moduli space.}
For our estimate, we choose the following distribution:
\begin{equation}
    P(c;q,q_0) = \alpha(q^2 - q^2_0)\frac{\text{e}^{\alpha(q^2 - q^2_0)c}}{\text{e}^{\alpha(q^2-q^2_0)}-1}
\label{eqn:prob_dist}
\end{equation}
where we define
\begin{equation}
    q_0 \equiv \sqrt{g_{a}^{2} + g_{b}^{2}}.
\end{equation}
This choice of probability distribution makes it exponentially difficult for states far out on the lattice to be light. This aligns with WGC intuition, according to which very massive string states should be viewed as black holes and, since we are modeling superextremal states only, their mass-to-charge ratio must therefore match the black hole extremality bound. In addition, for $\alpha = 1$ and large $q$, the mean of the above distribution is $1-1/ q^2$ which lines up with the mass-to-charge ratio of superextremal states of the heterotic string. The width of the distribution and the parameter $\alpha$ can then be thought of as resulting from symmetry breaking effects that contribute to particle masses. These effects are certainly less prominent for heavy states and this is captured by the decreasing width of the distribution as $q^2$ increases.

\begin{figure}[tp]
    \centering
    \includegraphics[width=0.54\textwidth]{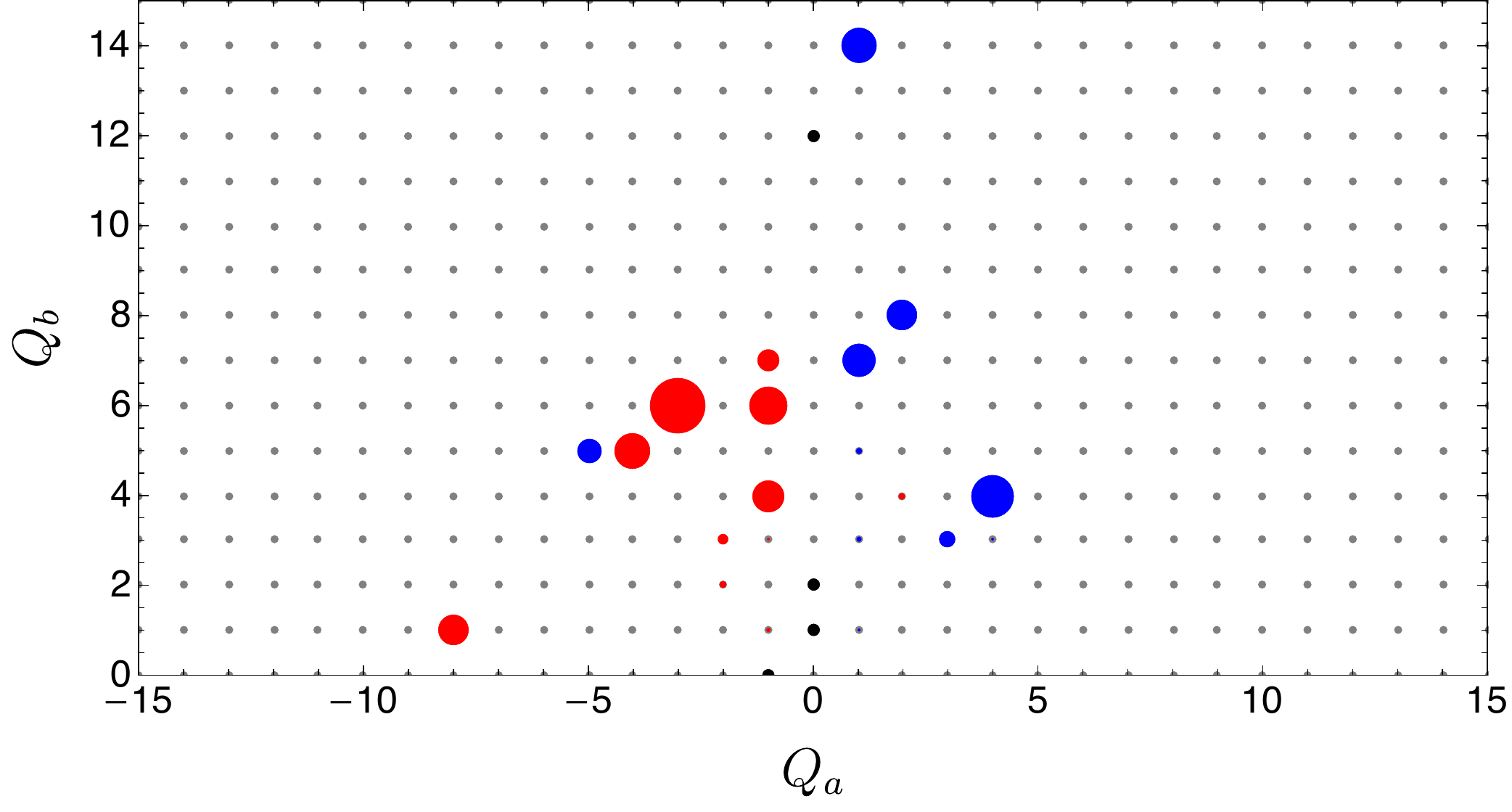}
    \includegraphics[width=0.45\textwidth]{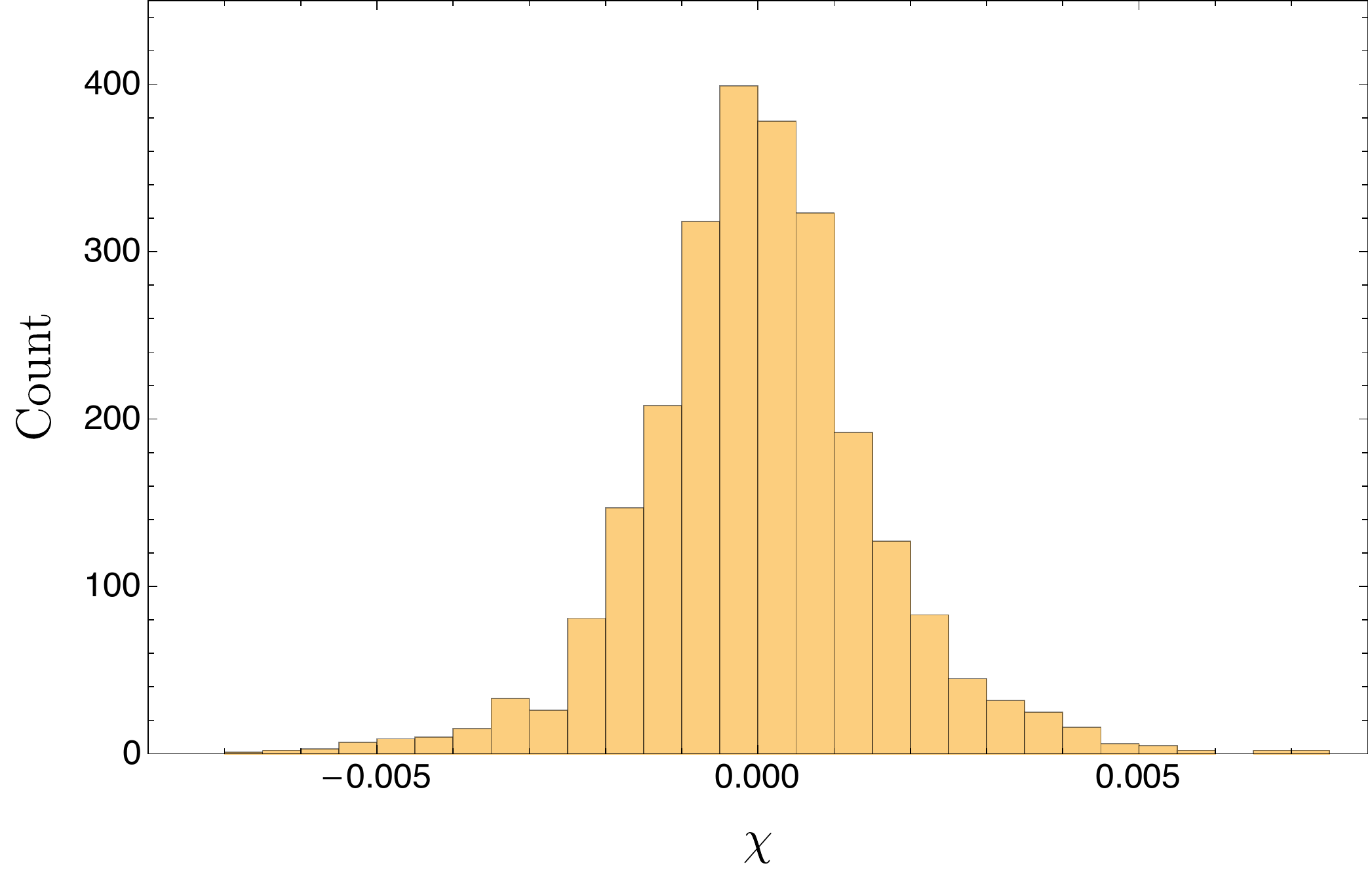}
    \caption{We show a sample charge lattice on the left. The relevant states are indicated by black (zero contribution), red (negative) and blue (positive) circles with the size of the circle being proportional to the magnitude. The right panel shows the distribution of kinetic mixing results. Due to our choice of PDF for the $c_{ij}$, we expect this distribution to be centered around $\chi_{ab} = 0$. Here we set $g_{a}$ = 0.1, $g_{b}$ = 0.2 and $\alpha$ = 6 and take $N=2500$ lattice realizations.
    }
    \label{fig:slwgc_estimate}
\end{figure}

Our choice of probability distribution is invariant under a $\mathbb{Z}_{2}$ symmetry where $Q_{a,b}\to -Q_{a,b}$ and this will result in a  distribution of $\chi$ that is centered at 0, as illustrated in the right panel of Figure~\ref{fig:slwgc_estimate}. In order to facilitate comparisons with string constructions in Section~\ref{sec:orbifolds}, we consider a particle spectrum which includes states with charges $( Q_a, \pm |Q_b|)$. As above, the masses of these states are different and are drawn at random from the distribution in Equation~\ref{eqn:prob_dist}. The contribution of any such pair of states is $\mu$-independent and given by:
\begin{align}
    \chi_{ij} = -\frac{g_a Q_a g_b |Q_b|}{48\pi^2}\ln \frac{m_{+}}{m_{-}} = -\frac{g_a Q_a g_b |Q_b|}{48\pi^2}\ln \frac{c_{+}}{c_{-}}
    \label{eqn:paircontribution}
\end{align}
where the subscripts on $c_{\pm}$ indicate the sign of the charge under $U(1)_b$ and we use the $(i,j)$ subscripts (with $j>0$, say) as a reminder that this is the contribution from one pair of lattice sites\footnote{Of course we can label the lattice by the coordinates $(Q_a, Q_b)$ but we use the notation $\chi_{ij}$ since it is simpler than $\chi_{Q_a,Q_b}$. The reader can always take $Q_a^{(i,j)} = i$ and $Q_b^{(i,j)} = j$.}. We can now derive an expression for probability distribution of $\chi_{ij}$ using the distribution we assumed for $P(c)$. We leave the details to Appendix~\ref{sec:appendix_ratio_dist} and quote the result here:
\begin{align}
    P(\chi_{ij}) =  \left|\frac{48\pi^2}{g_a Q_a g_b Q_b}\right| \times
    \begin{dcases}
        \frac{\ee^{\beta (1+r)}(1-\beta(1+r)) - 1}{(1-\ee^{\beta})^2 (1+r)(1+r^{-1})} & \text{for $r < 1$}\\
        \frac{\ee^{\beta (1+r^{-1})}(1-\beta (1+r^{-1})) - 1}{(1-\ee^{\beta})^2 (1+r)(1+r^{-1})} & \text{for $r > 1$}
    \end{dcases}
    \label{eqn:chi_dist}
\end{align}
where we have defined
\begin{align*}
    r \equiv \frac{c_+}{c_-} = \exp\left\{-\frac{48 \pi^2 \chi_{ij}}{g_a Q_a g_b |Q_b|}\right\} \quad \text{and} \quad \beta \equiv \alpha(q^2 - q_0^2).
\end{align*}
Note that under $\chi_{ij} \rightarrow -\chi_{ij}$, we have $r\rightarrow 1/r$ and the distribution is invariant implying $P(\chi_{ij}) = P(-\chi_{ij})$. The mean of $\chi$ then vanishes as a result of the vanishing of $\langle \chi_{ij}\rangle$ in agreement with the expectation outlined above. It is difficult to use $P(\chi_{ij})$ to determine the distribution of $\chi = \sum \chi_{ij}$ since the latter sum requires knowledge of which lattice sites represent states with masses below the species scale, as discussed in Section~\ref{sec:species_scale}. This question can only be answered by considering the full lattice realization at once. Thus, determining whether a particular (pair of) lattice sites contributes to the kinetic mixing is a complicated problem that depends on the probability distributions of all other lattice sites. For this reason, we approach this problem numerically and use the above analytic distribution to compute $\chi$ in the limit where the relevant lattice sites form a dense subset of the full charge lattice.

The numerical estimate of $\chi$ is carried out by creating an ensemble of lattice realizations and estimating the kinetic mixing by summing contributions of the form shown in Equation~\ref{eqn:paircontribution}. We show the $Q_b>0$ region of a sample charge lattice in the left panel of Figure~\ref{fig:slwgc_estimate}. The red and blue lattice sites correspond to states with masses below the species scale which are relevant for the estimate of kinetic mixing. The right panel of the same figure shows the distribution of $\chi$ in an ensemble of 2500 lattices. As explained above, the distribution is centered at $\chi =0$ but we generically expect a value of the mixing of order the standard deviation which is $\sigma\sim 10^{-3}$ in this case.

\begin{figure}[tp]
    \centering
    \includegraphics[width=0.45\textwidth]{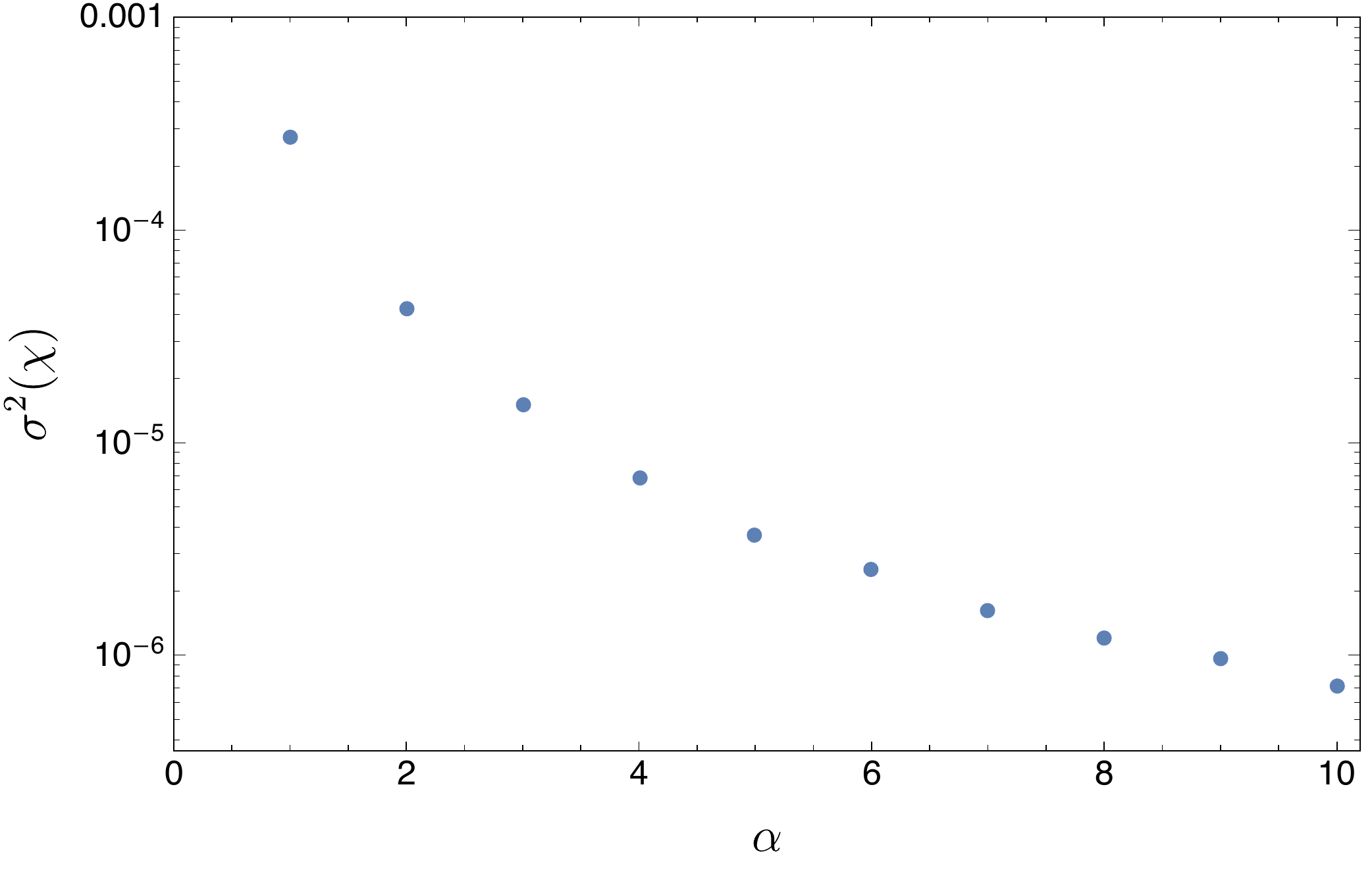}
    \includegraphics[width=0.45\textwidth]{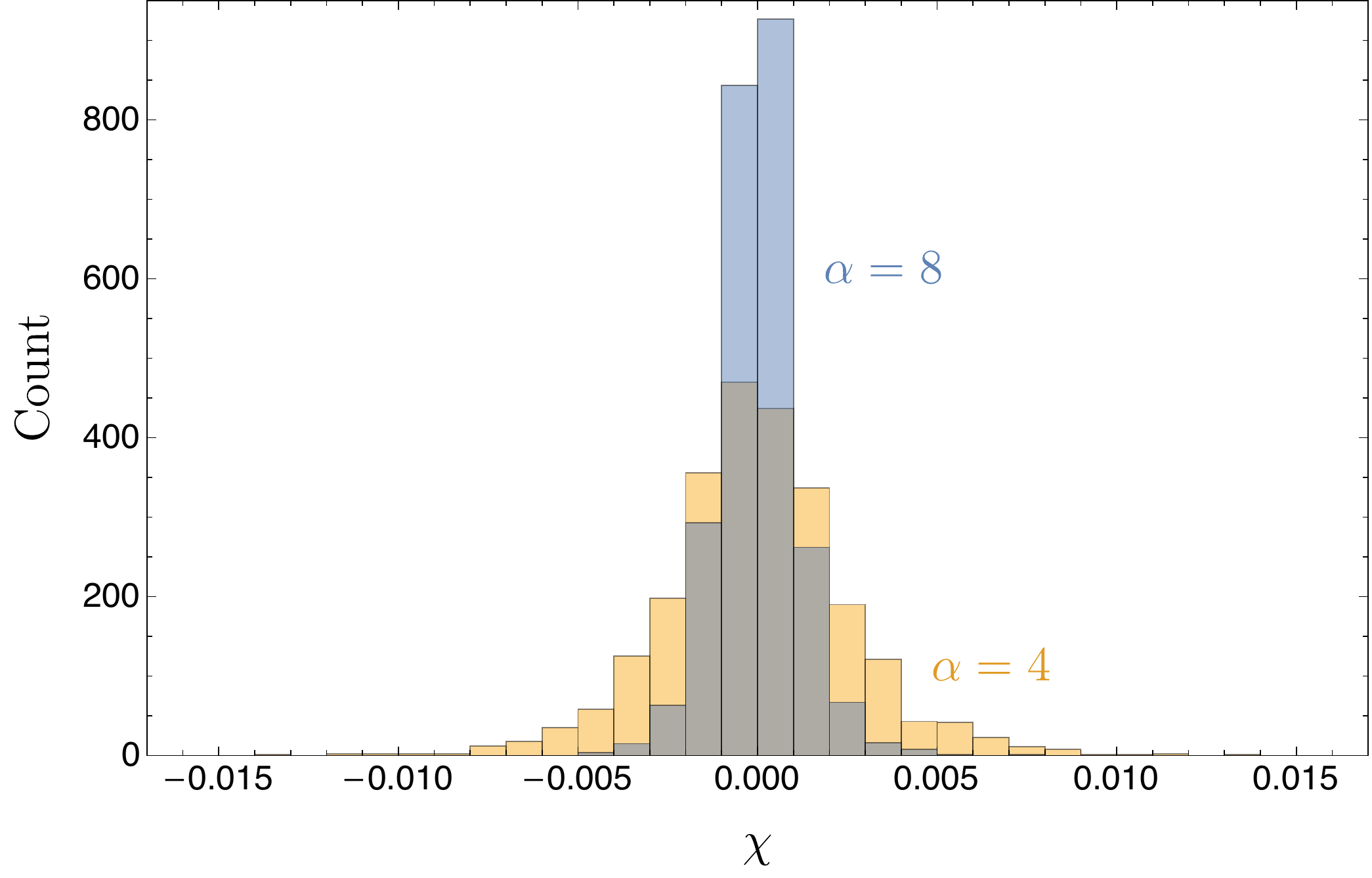}
    \caption{In the left panel, we show the variance as a function of $\alpha$. We fix $g_{a}$ = 0.1 and $g_{b}$ = 0.2. We notice that as $\alpha$ increases, the variance decreases. In the right panel, we show the distributions for $\chi_{ab}$ for $\alpha$ = 4 (yellow) and $\alpha$ = 8 (blue) and $N=2500$ lattice realizations.
    }
    \label{fig:var_alpha}
\end{figure}

\begin{figure}[tp]
    \centering
    \includegraphics[width=0.75\textwidth]{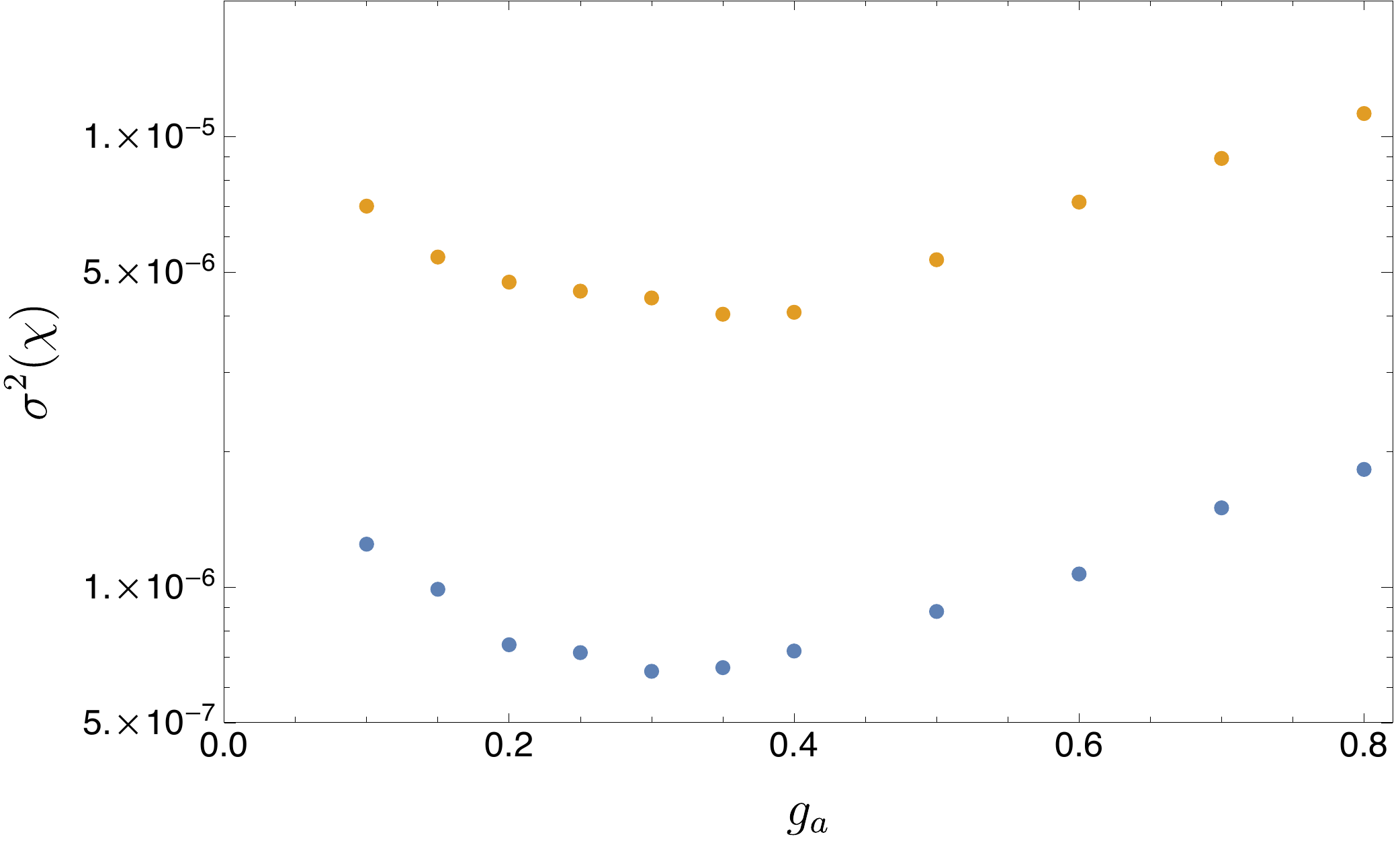}
    \caption{Here we show the dependence of the variance on the coupling. We fix $\alpha$ = 8. The blue curve shows $g_{b} = 0.2$ and the orange curve shows $g_{b} = 0.02$. We take $N=2500$ lattice realizations.}
    \label{fig:var_coupling}
\end{figure}

We now briefly comment on the dependence of $\chi$ on parameters of the distribution in Equation~\ref{eqn:prob_dist}. First, the parameter $\alpha$ controls how quickly the exponential rises. For larger $\alpha$, states approach the extremality bound at smaller charge and are on average more massive than the corresponding states with the same charge for a lower value of $\alpha$. As such, increasing $\alpha$ has the effect of discarding some states with large charges since these would now have masses that exceed the species scale. In turn, this decreases the value of the mixing on average and suppresses the tails of the $\chi$ distribution. We show this effect in Figure~\ref{fig:var_alpha}. Second, the coupling constants roughly control the eccentricity of the ellipse containing the relevant states. As such, increasing one of the couplings, say $g_a$, from a small value while holding the other fixed, again removes the contribution of some states with large charge and the kinetic mixing decreases. This continues until we have $g_{a}^{2}\sim g_{b}^{2} + \alpha^{-1}$ when the variance begins to increase again. This increase in variance can be attributed to the fact that we are limited by small numbers. For such large couplings, the lattice is populated by only a few superextremal states. Now if we make $g_{a}$ large, the most common configuration for these states is to lie on the $Q_{a}$ axis. The next most common configuration is when one state has a nonzero charge $Q_{a}$. This generates a larger contribution to the kinetic mixing compared to when two states have nonzero $Q_{a}$ charge, but partially cancel. As $g_{a}$ gets larger, having multiple states with nonzero $Q_{a}$ charge is exceedingly difficult and as a result the kinetic mixing, and its variance, both increase. We show this behavior in Figure~\ref{fig:var_coupling}.
Finally, we note that in the regime $g_{a} \gg g_{b}$, we found that most of the states have $Q_{a} = 0$. In the context of the Swampland, this amounts to lifting the $U(1)_{a}$ WGC towers above the species bound cutoff of $U(1)_{b}$. This effect is ubiquitous and we will see it again in Section~\ref{sec:5D_example}.

We now turn to a limit where we can use the analytical form of $P(\chi_{ij})$ to get an estimate of the kinetic mixing. Note that the probability distribution of $c = m/(qM_p)$ is controlled by the single parameter $\beta$. As such, in regions of the lattice where $\beta \ll 1$, we typically have a large number of light states below the species scale. By taking the limit of large $\alpha$ and small couplings $g_a, g_b \ll 1$ we can arrange for $\beta \ll 1$ in a region on the lattice that spans a large number of sites. In this limit, the relevant states for kinetic mixing can be approximated by a continuum inside the region $\beta \lesssim 1$. We can then calculate the variance of $\chi = \sum \chi_{ij}$ by noting that the $\chi_{ij}$ are independent variables (i.e. with a vanishing covariance). This allows us to first compute the variance of $\chi_{ij}$ and integrate the result over the ellipse defined by $\beta < 1$ to get the variance of $\chi$. Evaluating the integral numerically, we get:
\begin{align}
    \langle \chi_{ij}^2 \rangle = 1.49\times \left(\frac{g_a g_b Q_a Q_b}{48 \pi^2}\right)^2 \approx \frac32  \left(\frac{g_a g_b Q_a Q_b}{48 \pi^2}\right)^2.
    \label{eqn:chiij_variance}
\end{align}
Finally, summing the contributions from the $\beta < 1$ half-ellipse, we get:
\begin{align}
    \langle \chi^2 \rangle = \frac{(1+ \alpha(g_a^2 + g_b^2))^3}{73728 \pi^3 \alpha^3 g_a g_b}.
    \label{eqn:chi_variance}
\end{align}
In its regime of applicability, this is a decreasing function of each of its three parameters as expected from the explanation above. For example, for $\alpha = 10^3$ and $g_a = g_b = 10^{-3}$, we get $\chi \sim 10^{-5}$. Surprisingly, this also gives a good estimate for the benchmark values of couplings and $\alpha$ shown in Figure~\ref{fig:slwgc_estimate}.

\section{Computing $\chi$ in Explicit Examples}
\label{sec:constructions}

In this section we consider explicit QFT and string theory examples and compute the kinetic mixing parameter in these constructions. We aim to support the above arguments based on the sLWGC by showing that generic backgrounds typically lead to mixing between Abelian gauge groups in the low-energy theory. First we consider a 5D $U(1)$ gauge theory compactified on a circle. The associated graviphoton provides us with the second $U(1)$. We further consider scalars in this 5D theory which satisfy the LWGC. Upon compactification on $S^{1}$, we get a 4D theory with associated KK towers for each scalar that are charged under both gauge groups. These states generate kinetic mixing at 1-loop provided we turn on a background Wilson line. The next class of examples we consider are toroidal orbifold compactifications of the $E_8\times E_8$ heterotic string. In particular, we consider the ensemble of $\mathbb{Z}_{6-{\rm II}}$ orbifolds of~\cite{Nilles:2014owa} and compute the kinetic mixing distribution of these 1858 MSSM-like models. Finally, we work out the kinetic mixing in Type IIB supergravity compactified on the mirror quintic manifold. Here we perform an area average over the complex structure moduli space to produce an estimate of the kinetic mixing.

\subsection{5D $U(1)$ Example}
\label{sec:5D_example}
\begin{figure}[tp]
    \centering
    \includegraphics[width=\textwidth]{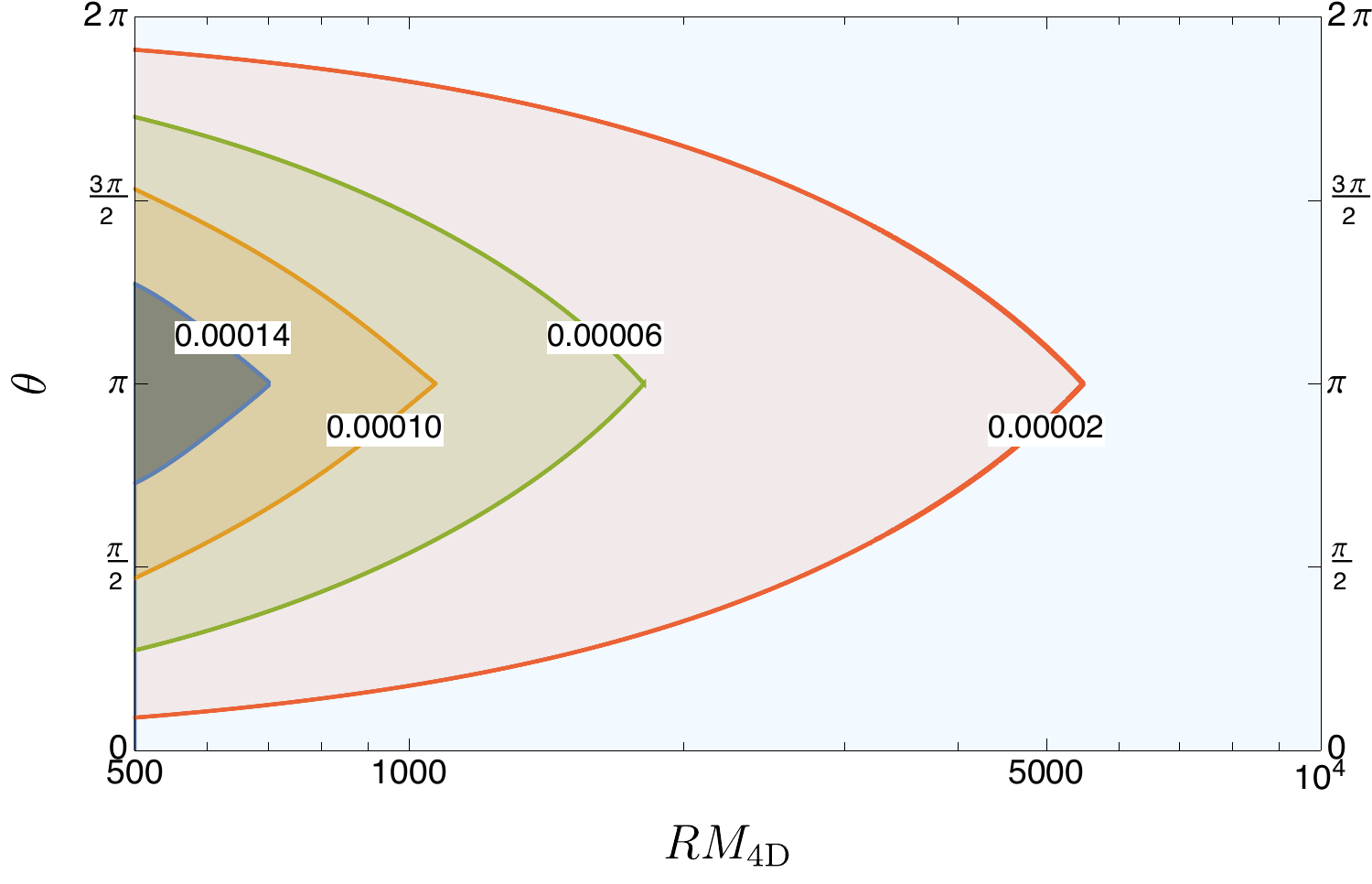}
    \caption{Here we show the one-loop $\chi$ generated from integrating out the KK modes of the graviton, photon and scalars as a function of the compactification radius $R$ and the Wilson line $\theta$. We fix $e_{\text{4D}} = 10^{-3}$. $\theta$ induces a mass splitting between oppositely charged states leading to the increase in $\chi_{ab}$ as a function of $\theta$. The symmetries of the theory, namely parity and large gauge transformations in the compact dimension, imply that the theory is identical at $\theta = \pi \pm \alpha$. We discuss this in more detail in the text. Increasing $R$ decreases $e_{\text{KK}}$ which eventually lifts all states charged under both U(1)s above the species scale, thereby decreasing $\chi$.}
    \label{fig:wilson_line_estimate}
\end{figure}

We start with the simplest example that illustrates mixing from a lattice by considering a 5D Abelian gauge theory coupled to gravity. The LWGC implies that each site on the charge lattice $\Gamma$ is populated with a state.\footnote{Assuming a sublattice won't qualitatively change the results and is similar to rescaling the charges of the states by an $\mathcal{O}$(1) number.} We will take these states to be represented by complex scalars $\phi_i$ of mass $m_i$ and charge $q_i$.
This is given by the following action:
\begin{equation}
    S = \int d^{5}x\sqrt{-g}\Bigg[\frac{1}{2\kappa_{\rm 5D}^{2}}\mathcal{R} - \frac{1}{4e^{2}_{5\text{D}}}F_{MN}F^{MN} - \sum_{i\in\Gamma}(D_{M}\phi_{i})^{\dagger}(D^{M}\phi_{i}) - m^{2}_{i}|\phi_{i}|^{2}\Bigg].
\end{equation}
For the following discussion, we turn off the dilaton coupling (i.e. $\alpha = 0$ in Equation~\ref{eqn:gen_WGC}). For a 1-form gauge field in 5D, this implies that a state of charge $q$ has a mass satisfying
\begin{equation}
    \frac{2}{3}m_{\rm 5D}^{2} \leq e^{2}_{5\text{D}}q^{2}M^{3}_{5\text{D}} \quad \rightarrow \quad m_{\rm 5D}^{2} \leq \frac{3}{2}e^{2}_{4\text{D}}q^{2}M^{2}_{4\text{D}}
\label{eqn:WGC_5D_mass}
\end{equation}
We will take each of our scalars to have a mass saturating the bound in Equation~\ref{eqn:WGC_5D_mass}.
This is conservative because of two effects. The first is that the contribution of states with $\pm n$ is proportional to $\log\left(m_+/m_-\right)$ and the fractional difference in the masses decreases as $m_{\rm 5D}$ increases. The second effect is a reduction in the number of contributing particles since the masses of heavy states saturate the species bound for a smaller total number of states.
We compactify on a circle of radius $R$ yielding a $U(1)_{\text{F}}\times U(1)_{\text{KK}}$ gauge theory where $U(1)_{\text{F}}$ descends from 5D with coupling $e_{4\text{D}}$ in 4D and $U(1)_{\text{KK}}$ arises from the graviphoton with coupling $e^{2}_{\text{KK}}R^{2} = 16\pi G$.\footnote{The 5D and 4D gauge coupling and Planck mass $M_{d}$ are related via $e^{2}_{d} = 2\pi R e^{2}_{d-1}$ and $M_{d-1}^{d-3} = 2\pi RM_{d}^{d-2}$.} A KK tower is generated for each scalar. Furthermore, under the assumption of genericity, there can exist a nonzero Wilson line $\theta = \int dy A^{5}$ along the compact direction which shifts the mass of each state.
The 4D theory then contains a full lattice of states. Each lattice site is labelled by $(q,n)$ and is populated by a state with mass given by
\begin{equation}
    m_{\rm 4D}^{2} = m_{\rm 5D}^{2} + \frac{1}{R^{2}}\Bigg(n - \frac{q\theta}{2\pi}\Bigg)^{2}
\label{eqn:scalar_mass}
\end{equation}
The charge under $U(1)_{\text{F}}$ is denoted by $q$ and $n\in\mathbb{Z}$ denotes the charge under $U(1)_{\text{KK}}$. In addition to the scalars, the KK compactification generates a tower for the graviton as well as the photon. These states are charged under $U(1)_{\text{KK}}$, but uncharged under $U(1)_{\text{F}}$. Their masses are given by setting $q = 0$ in Equation~\ref{eqn:scalar_mass}. Although these states don't contribute to the mixing, they are important because they are light states included in our effective theory which determine the species scale. We show our results in Figure~\ref{fig:wilson_line_estimate}. We fix the 4D gauge coupling to $e_{\text{4D}} = 10^{-3}$.
This leaves the compactification radius $R$ and the Wilson line $\theta$ as the two free parameters.

The parameter $R$ controls the size of the extra dimension, and in turn also controls the size of $e_{\text{KK}}$. Increasing $R$ decreases the KK gauge coupling. As we begin to develop a hierarchy between the two gauge couplings in our 4D theory, we also begin to lift the tower of one of our $U(1)$'s above the species bound cutoff of the other U(1). This means that the species bound of our theory is saturated by states which are uncharged under one of the U(1) gauge groups. This can be seen in Figure~\ref{fig:wilson_line_estimate}, where increasing $R$ leads to a decreased magnitude of kinetic mixing.

The parameter $\theta$ controls the breaking of an exact $\mathbb{Z}_{2}$ symmetry.\footnote{This is a global symmetry which we expect to either be broken or gauged. When $\theta \neq 0$ mod $2\pi$, it is broken. We relegate discussion of the gauged case to Section~\ref{sec:loopholes}.} When $\theta = 0$, states with charge $(\pm q, n)$ are exactly degenerate as can be seen from Equation~\ref{eqn:scalar_mass}. In this limit, the contribution to the kinetic mixing, as shown in Equation~\ref{eqn:kinmix_formula}, from both of these states exactly cancels.
As we increase $\theta$, we increase the mass splitting between these oppositely charged states. This allows for a nonzero contribution to the mixing. This is reflected in Figure~\ref{fig:wilson_line_estimate}, where larger values of $\theta$ correspond to an increased magnitude of kinetic mixing.
We note that the mixing increases until $\theta = \pi$ and then decreases. In particular, the mixing is symmetric about $\theta = \pi$.
We can consider the symmetries of our theory to better understand this behavior. The theory is invariant under $\theta \to \theta + 2\pi$ due to large gauge transformations in the compact dimension. This is consistent with the fact that the mixing vanishes at $\theta = 0$ and $\theta = 2\pi$. Furthermore, parity in the fifth dimension implies that $\theta \to -\theta$ also leaves the theory invariant. Combining these two symmetries, we find that the theory is identical at $\theta = \pi \pm \alpha$, and hence the mixing contours are symmetric about $\theta = \pi$.
Since our theory is endowed with an integer charge lattice, these symmetries can also be understood using GL(2,$\mathbb{Z}$) transformations. Suppose we define the generators of GL(2,$\mathbb{Z}$) as follows
\begin{equation}
    S = \begin{pmatrix}
    0 & -1 \\
    1 & 0
    \end{pmatrix} \qquad
    T = \begin{pmatrix}
    1 & -1 \\
    0 & 1
    \end{pmatrix} \qquad
    P = \begin{pmatrix}
    1 & 0 \\
    0 & -1
    \end{pmatrix}
\end{equation}
According to these definitions, $\theta \to \theta + 2\pi$ is given by $\mathcal{L}_{\theta}(T\vec{a}) = \mathcal{L}_{\theta + 2\pi}(\vec{a})$ and $\theta \to -\theta$ is given by $\mathcal{L}_{\theta}(P\vec{a}) = \mathcal{L}_{-\theta}(\vec{a})$, where $\vec{a}$ is a vector containing our two gauge fields.
As an example, suppose we consider our theory at $\theta_{1} = \pi/2$ and $\theta_{2} = 3\pi/2$. For a state with $q = 1$ under $U(1)_{\rm F}$, the lightest state, given by Equation~\ref{eqn:scalar_mass}, corresponds to $n = 0$ for $\theta_{1}$ and $n = 1$ for $\theta_{2}$. We can either fix the gauge eigenbasis, in which case the lightest mass eigenstate has different charges at the two values of $\theta$, or we can fix the charges of the lightest mass eigenstate, and redefine our gauge eigenbasis. We choose to do the latter. Performing this gauge basis change, we find that the spectrum of the theory is identical at $\theta_{1}$ and $\theta_{2}$ producing identical mixings, consistent with our discussion of the symmetries above.

\subsection{Heterotic string theory on orbifolds}
\label{sec:orbifolds}
Kinetic mixing has been studied in the three popular settings for string phenomenology: heterotic string compactifications, brane-world scenarios for the Type II string and F-theory. The earliest investigations were based on the heterotic string but later studies also found mixing in Type II constructions~\cite{Abel:2008ai} as well as F-theory compactifications~\cite{DelZotto:2016fju}. In this section, we will focus on orbifold compactifications of the heterotic string as an example where the generic expectation for kinetic mixing can be estimated. We will study the statistics of the mixing parameter in a large sample of semi-realistic heterotic orbifold models. Our aim is to compare these estimates with the phenomenological ones derived above.

In heterotic orbifolds, being exact CFT constructions, the full spectrum of states is known which allows for an explicit calculation of threshold corrections to kinetic mixing. This is similar to the computation of threshold corrections to gauge coupling constants which have been studied in~\cite{Kaplunovsky:1987rp}. The latter formalism was extended to compute kinetic mixing between the gauge bosons of two different U(1)'s in~\cite{Dienes:1996zr} and we now review the relevant details. If we take $\chi = 0$ at the string scale, then the low-energy kinetic mixing between $U(1)_a$ and $U(1)_b$ at one loop is given by:
\begin{equation}
    \left\{\frac{\chi}{g_a g_b}\right\}(\mu) = \frac{b_{ab}}{16\pi^2}\ln\frac{M^2_{\rm st}}{\mu^2} + \frac{1}{16\pi^2}\Delta_{ab}
\end{equation}
where $M_{\rm st}$ is the string scale. In the above expression, the first term describes the running of the mixing parameter due to the presence of light bi-charged matter and the second term gives the string threshold correction to kinetic mixing due to the presence of massive states above the string scale. We are mainly interested in estimating the magnitude of the kinetic mixing parameter and so we will focus on the value of the threshold correction $\Delta_{ab}$ and will ignore contributions of the first term. That said, the threshold correction is computed by:
\begin{equation}
    \Delta_{ab} =
    \int_F \frac{\dd^2 \tau}{\tau_2}\left(B_{ab}(\tau, \bar{\tau}) - b_{ab}\right)
\end{equation}
where we have ignored a universal term proportional to $k_{ab}$, the coefficient of the $1/z^2$ pole in the OPE $J_a(z)J_b(0)$ between the $U(1)$ worldsheet currents. This is justified since we have chosen the two $U(1)$ generators to be orthogonal in the UV theory implying the vanishing of the $1/z^2$ pole in the OPE of their worldsheet currents. What remains then is an integral over the $PSL(2,\mathbb{Z})$ fundamental domain of
\begin{equation}
    B_{ab}(\tau,\bar{\tau}) \equiv |\eta(\tau)|^{-4} \sum_{{\rm even }\;{\bf s}}(-1)^{s_1 + s_2} \frac{\dd Z_\Psi({\bf s}, \bar{\tau})}{2\pi i \dd \bar{\tau}} {\rm Tr}_{s_1}\left( Q_a Q_b (-1)^{s_2 F} q^H \bar{q}^{\bar{H}}\right)
\end{equation}
from which we have subtracted the contribution of the massless states, $b_{ab}$. In the above expression, $\eta(\tau)$ is the Dedekind eta function and $Z_\Psi({\bf s}, \bar{\tau})$ is the partition function of the right-moving non-compact complex fermion with spin structure $\mathbf{s}$. The threshold correction $\Delta_{ab}$ then calculates the effect of integrating out massive string states above the string threshold.

After developing the above formalism, \cite{Dienes:1996zr} applied it to three standard-like models~\cite{Antoniadis:1990hb,Faraggi:1991jr,Faraggi:1993sr} as an example. The kinetic mixing in all these models surprisingly vanishes but this result is not robust to $\emph{any}$ correction to the mass spectrum and the expectation is that kinetic mixing is still present from other low-energy effects.
Subsequently,~\cite{Goodsell:2011wn} investigated mixing in symmetric Abelian factorizable orbifolds arguing, along the lines of~\cite{Kaplunovsky:1987rp,Dixon:1990pc}, that models with non-zero mixing are ones that have an $\mathcal{N}=2$ subsector. For this class of models, the existence of an $\mathcal{N}=2$ subsector requires an orbifold point group of non-prime order so that certain twists fix one of the compact tori leaving additional unbroken SUSY charges.

We will focus on these $\mathcal{N}=2$ subsectors but briefly comment on the other sectors. The orbifold models we are considering also have $\mathcal{N}=1$ (e.g. the first twisted sector) and $\mathcal{N}=4$ (the untwisted sector) subsectors but these do not provide moduli-dependent contributions to the kinetic mixing and can be ignored for our purposes as we now briefly explain. We begin by considering the untwisted sector. In this sector it is easy to see that the spin structure dependent part of $B_{ab}$ becomes:
\begin{equation}
    \sum_{{\rm even}\;{\bf s}}(-1)^{s_1 + s_2} \frac{\dd Z_\Psi}{\dd \bar\tau} \times Z_\Psi^3
    = \frac{1}{4}\frac{\dd}{\dd\bar{\tau}}\sum_{{\rm even}\;{\bf s}}(-1)^{s_1 + s_2} Z_\Psi^4,
\end{equation}
which vanishes due to Jacobi's abstruse identity. Turning to the $\mathcal{N}=1$ subsectors
it is easy to see that any contribution they provide
cannot depend on the K\"{a}hler and complex structure moduli $T_i$ and $U_i$ describing the compact torus since such states reside at fixed points and cannot probe the torus geometry. Any contribution to $\Delta_{ab}$ must then be an additive constant. However, since the contribution from the $\mathcal{N}=2$ sector is moduli dependent and we will choose rough $\mathcal{O}(1)$ numbers for these moduli, computing the $\mathcal{N}=1$ contributions is of little value for our estimate.\footnote{For a more accurate calculation, one could average over moduli space and produce an area weighted probability distribution but we leave this for future work.}

The focus is then on $\mathcal{N} = 2$ subsectors which are twisted sectors whose twist fixes one of the three directions of the compact torus. String states in these sectors are not localized along the fixed direction and can probe the torus geometry allowing for dependence of the threshold correction on the torus moduli. The form of the moduli dependence was first calculated in~\cite{Kaplunovsky:1987rp,Dixon:1990pc} for threshold corrections to gauge couplings and then adapted to the kinetic mixing scenario in~\cite{Goodsell:2011wn}. The exact dependence will not be important for our estimates. Instead, with $\mathcal{O}(1)$ values for the moduli vevs, we have:
\begin{equation}
    \Delta_{ab} = \mathcal{O}(1)\times \sum_i \frac{b_{ab}^i |G^i|}{16\pi^2 |G|}
\end{equation}
where $b_{ab}^i$ is evaluated from the massless states in the $\mathcal{N}=2$ subsector that fixes the $i$-th torus, $|G|$ is the order of the orbifold group and $|G^i|$ is the order of its subgroup fixing the $i$-th torus.

\begin{figure}[tp]
    \centering
    \includegraphics[width=0.75\textwidth]{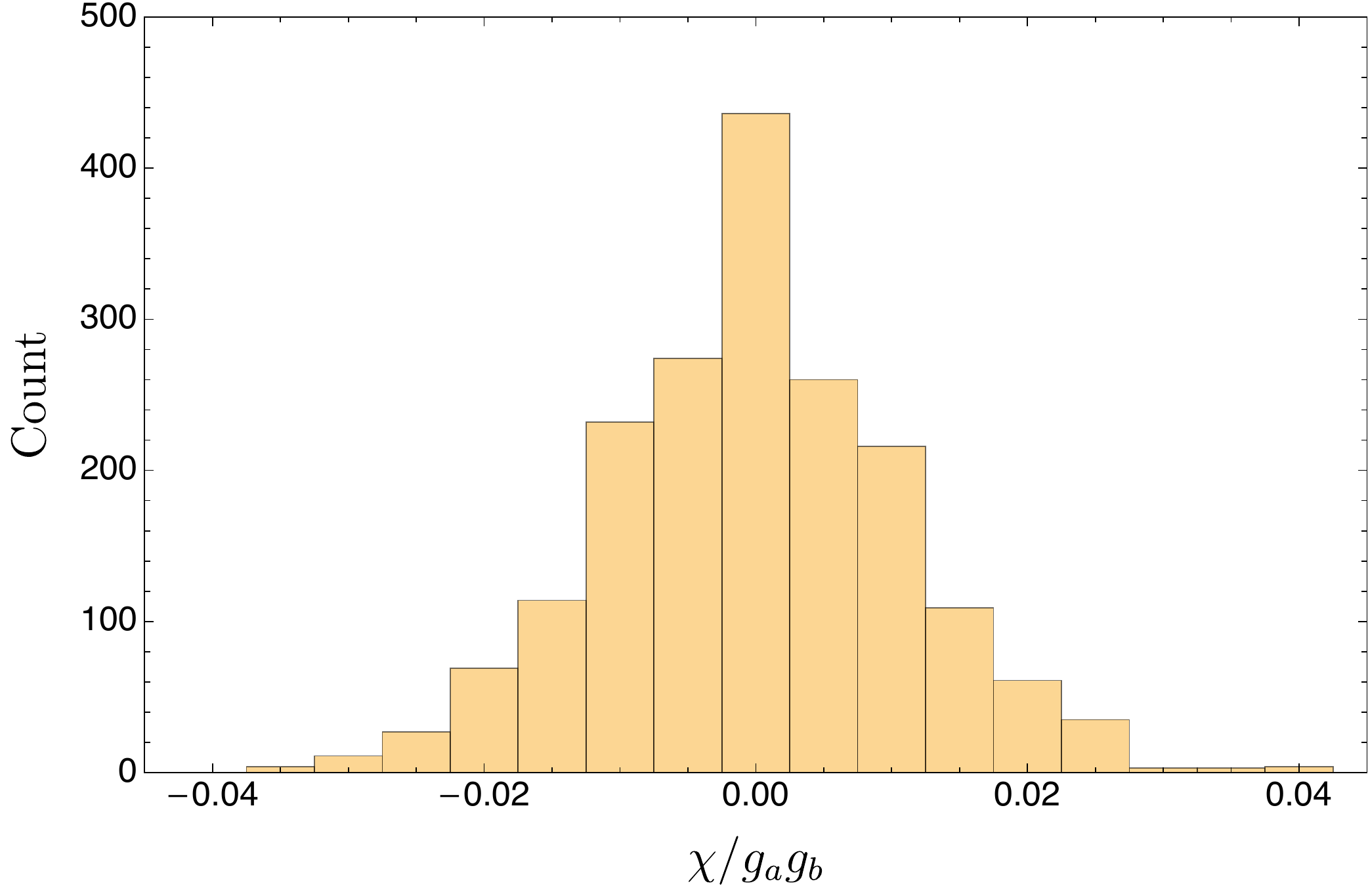}
    \caption{We show the distribution of the kinetic mixing coefficient in a sample of 1858 MSSM-like $\mathbb{Z}_{6-{\rm II}}$ orbifolds.
    }
    \label{fig:orbifold_distribution}
\end{figure}

Using this expression for the mixing, we evaluate $\chi/g_a g_b$ for models from the mini-Landscape~\cite{Nilles:2014owa} which contains a large number of inequivalent heterotic orbifold models. The results are shown in Figure~\ref{fig:orbifold_distribution}. This is similar to what we saw in our phenomenological estimate: the distribution again has zero mean and a variance that is comparable in magnitude to the distributions of Section~\ref{sec:sLWGC_stat}. This gives us confidence that the phenomenological estimate carried out above reproduces the generic expectation in the string Landscape. In addition, it shows that modeling the superextremal states is sufficient for our estimate given the agreement with the string computation which takes into account all charged states.

\subsection{Type II on a Calabi-Yau Manifold}
\label{sec:quintic}
In the previous section we considered models based on the heterotic string on orbifolds (CFT constructions) but we can also estimate the mixing using supergravity, i.e. using the low energy effective description of string theory. For concreteness we will take an example based on Type IIB supergravity compactified to four dimensions on the mirror quintic where the mixing is determined in terms of the one-dimensional complex structure moduli space and thus receives no $\alpha'$ corrections. In addition, we imagine we are at a fixed point in K\"{a}hler moduli space at large volume so that the supergravity approximation is valid. The supergravity approach can be used to estimate the kinetic mixing distribution on backgrounds more general than orbifolds, such as flux compactifications (see for example~\cite{Abel:2008ai}).

We begin by recalling the bosonic terms in the Type IIB low-energy action in Einstein frame (see for example~\cite{Polchinski:1998rr}):
\begin{align}
    2\kappa_{10}^2 S^{{\rm IIB}}_{10D} = \int \dd^{10} x \sqrt{-G}\left(
    R - \frac12 \pd_\mu \phi \pd^\mu \phi - \frac12 \ee^{-\phi} |H_3|^2 - \frac12 \ee^{2\phi} |F_1|^2 - \frac12 \ee^{\phi}  |F_3|^2 - \frac14 |F_5|^2
    \right)
\end{align}
where
\begin{align}
    &H_3 = \dd B_2, \qquad F_1 = \dd C_0, \qquad F_3 = \dd C_2 - C_0 \wedge H_3\\
    &F_5 = \dd (C_4 - \frac12 C_2 \wedge B_2)\equiv \dd \tilde{C}_4 \quad \text{with} \quad F_5 = \star F_5.
\end{align}
Dimensional reduction of the above action on a CY threefold leads to an effective action in 4D whose field content is determined by topological data (namely Hodge numbers $h^{p,q}$) of the compact space. In particular, the RR potentials decompose into a sum over a basis of harmonic forms on the CY where the coefficients are 4D massless fields. Of relevance to the gauge kinetic function is the real symplectic basis of 3-forms $(\alpha_i, \beta^j)$ with $i = 0, \ldots, h^{2,1}$. The expansion of $\tilde{C}_4$ thus includes $2(h^{2,1} + 1)$ 1-forms in 4D (i.e. gauge potentials which are the coefficients of $(\alpha_i, \beta_j)$). However, the self-duality condition imposed on $F_5$ halves the number of 4D gauge fields. In total we therefore get $h^{2,1} + 1$ gauge fields, one of which resides in the 4D $\mathcal{N} = 2$ gravity multiplet and the remaining $h^{2,1}$ are part of the vector multiplets. In 4D, the terms relevant for our study of these gauge potentials are given by:
\begin{align}
    S^{{\rm IIB}}_{4D} \supset \int \dd^4 x\sqrt{-g} \left(
    \frac{1}{8\pi} {\rm Im} {\mathcal M}_{ij} F^i_{\mu\nu} F^{j\mu\nu}
    + \frac{1}{8\pi} {\rm Re} {\mathcal M}_{ij} F^i_{\mu\nu}\tilde{F}^{j\mu\nu}
    \right)
\end{align}
where the gauge kinetic function can be derived from the periods of the holomorphic 3-form $\Omega$ on the CY and depends on the complex structure parameters. Explicity, we have:
\begin{align}
    &z^i = \int_{\rm CY} \Omega \wedge \beta^i, \qquad \mathcal{G}_i = \int_{\rm CY} \Omega \wedge \alpha_i, \label{eq:periods} \\
    &\mathcal{M}_{ij} =
    \overline{\mathcal{G}}_{ij} + 2i \frac{{\rm Im}\mathcal{G}_{im}z^m {\rm Im}\mathcal{G}_{jn} z^n}{{\rm Im}\mathcal{G}_{kl} z^k z^l } \label{eq:Mmatrix}
\end{align}
where $\mathcal{G}_{ij} \equiv \pd_{z^i} \mathcal{G}_j$, recalling that the periods are not independent and that one may regard $\mathcal{G}_i(z)$ as functions of $z^i$ which are homogeneous coordinates on the complex structure moduli space. More details on the effective theory of Type II supergravity can be found in~\cite{Grimm:2005fa}.

\begin{figure}[tp]
    \centering
    \includegraphics[width=0.79\textwidth,valign=m]{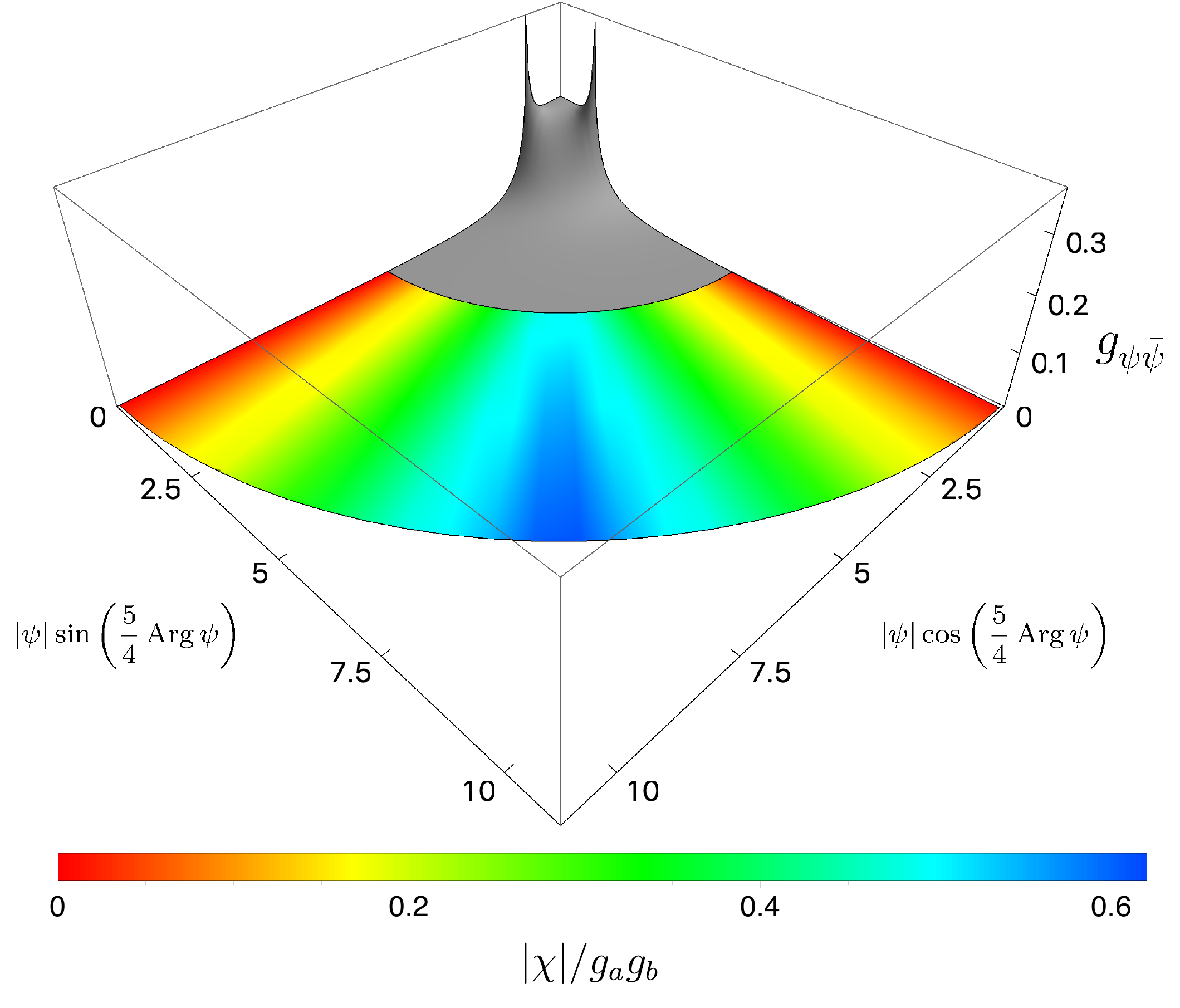}
    \caption{We show the moduli space metric $g_{\psi\bar{\psi}}$ as a function of the complex parameter $\psi$ in the range $0\leq {\rm Arg}[\psi] < 2\pi/5$. The color coding corresponds to the size of the kinetic mixing coefficient at that point.}
    \label{fig:quintic_estimate}
\end{figure}

As an example CY$_3$, we consider the mirror of the quintic hypersurface in $\mathbb{CP}^4$ which has $h^{2,1} = 1$ and is thus characterized by a single complex structure parameter we call $\psi$. The periods and the geometry of the moduli space have been studied long ago in~\cite{Candelas:1990rm} and the results can be expressed analytically in terms of hypergeometric functions. These periods can be used to determine the mixing between the two photons in the 4D theory according to the gauge kinetic function described above - the two photons being the graviphoton and the additional one in the vector multiplet. We show the metric and the kinetic mixing in the $\psi$-plane in Figure~\ref{fig:quintic_estimate}.

At this point, it is necessary to make a few comments. First, the choice of basis 3-forms $(\alpha_i, \beta^j)$ is not unique and any other basis related to the one chosen by a symplectic transformation is equally valid. In the 4D theory, this corresponds to a choice of electric-magnetic duality frame. While all these choices are physically equivalent, they do not always allow for a weakly coupled description and one must be careful when extracting information about kinetic mixing. In addition, the kinetic mixing, like the couplings, depends on the choice of duality frame but one ideally wants a basis-independent prediction for the mixing. In order to deal with these issues, we adopt a specific basis choice given by placing all $\mathcal{M}_{ij}$ matrices in the so-called genus 2 Siegel's fundamental domain\footnote{Technically, we work with $\mathcal{M}^*_{ij}$ since this has positive definite imaginary part but this distinction does not affect the results.}. We describe Siegel's fundamental domain and the procedure we follow in Appendix~\ref{app:siegel}. Despite these duality transformations, there remain parts of moduli space (colored gray in Figure~\ref{fig:quintic_estimate}), where there is no weakly coupled description and we excise this region. Finally, since the ${\rm Arg}[\psi] = \{0,2\pi/5\}$ rays are identified, we see a monodromy effect like that discussed in Section~\ref{sec:5D_example}. The quintic provides an example where the $\chi$ distribution has a nonzero mean. This shows that the phenomenology of kinetic mixing could be much richer than one might expect from the simplest estimates shown in previous sections.

\section{Loopholes}
\label{sec:loopholes}
In this section we review potential loopholes that could obstruct the above genericity arguments and lead to vanishing kinetic mixing between low energy $U(1)$'s. These loopholes typically correspond to finely tuned regions of parameter space and/or enjoy an enhanced symmetry.

\begin{itemize}
    \item {\bf Non-Abelian unification:}
    The simplest loophole to the above genericity arguments is the presence of a symmetry between charged particles running in the loop that leads to cancellations when evaluating the total kinetic mixing from all available species. These symmetries are easy to come by since one can consider, for instance, a non-Abelian $SU(N)$ gauge group that is Higgsed down to a product of $U(1)$ factors by going on the Coulomb branch. Since the generators of $SU(N)$ are traceless, each $SU(N)$ representation decomposes into $U(1)$ representations where particles have charges with a vanishing sum under each $U(1)$. It is then easy to see that kinetic mixing from such a spectrum would vanish (e.g. by considering all the particles with charge $q_*$ under the first $U(1)$). In this scenario, however, Higgs insertions can still allow for a kinetic mixing effect (albeit a mass suppressed one). Estimates of this effect have been carried out in the context of GUT models (see for example~\cite{Gherghetta:2019coi} for a recent discussion).

    \item {\bf Charge conjugation in the dark sector:}
    This is similar to the unification scenario discussed above and implements a symmetry in the spectrum that leads to cancellations preventing loop corrections from generating kinetic mixing. We saw this explicitly in the example in Section~\ref{sec:5D_example}, where $\theta = 0$ corresponds precisely to the case where the 4D particle spectrum has a charge conjugation symmetry. Alternatively, one can note that the operator $F^{(a)}_{\mu\nu}F^{(b)\mu\nu}$ is not invariant under separate charge conjugations and is thus forbidden if the symmetry is left intact. In a theory of quantum gravity, this charge conjugation must be a discrete gauge symmetry such as the one obtained by Higgsing $SO(3)$ with a Higgs in the 5-dimensional irreducible representation to $U(1) \rtimes \mathbb{Z}_2 \simeq O(2)$. We see then that in a theory with strict $U(1)$ (rather than $O(2)$) gauge groups, this charge conjugation is absent. In addition, in cases where it is present, there are additional physical effects that could potentially distinguish the two gauge groups such as `Alice' strings.

    \item {\bf Fine-tuned loci in moduli space:}
    In theories with supersymmetry, there could be a moduli space of vacua parametrized by the vev of scalar fields. In these cases, it might be possible to tune the moduli to a value that leads to vanishing kinetic mixing. However, such loci are typically lower dimensional submanifolds of the moduli space and would thus be missed by generic field vevs. An example is an $\mathcal{N}=2$ $SU(N)$ gauge theory where one can give a vev to the scalar in the vector multiplet to Higgs $SU(N) \rightarrow U(1)^{N-1}$. The low-energy Lagrangian then contains a gauge coupling matrix that depends on the particular vev configuration; $\mathcal{L} \supset \int d^2\theta \tau^{ij}(\phi) W_i W_j$. However, the mixing in the case would vanish only on a lower dimensional subspace of the full moduli space and is non-zero at a generic point.

    \item {\bf Prime orbifolds:}
    This scenario is similar to the previous case since orbifolds are special points in moduli space. In particular, prime orbifolds provide many examples where kinetic mixing vanishes. In this case, generic points on moduli space describe orbifolds where the singularities have been blown up by giving vevs to twisted moduli. This alters the spectrum of states and is expected to generate kinetic mixing. In addition, these prime orbifold examples typically include massless bi-charged matter which must be lifted for the model to become phenomenologically viable. Giving these states mass would again generically induce a kinetic mixing signal.

    \item {\bf Braneworld Scenarios:}
    In this case, the two $U(1)$ gauge groups can be localized on separate branes and the kinetic mixing between them could be exponentially suppressed if it relies only on the overlap between the two wavefunctions localized on each brane. In the presence of light bi-charged bulk modes, kinetic mixing can still be appreciable. An example is considered in~\cite{Abel:2008ai}, where a bulk $B_{\mu\nu}$ field mediates the interaction and leads to a wide range of values for the mixing parameter depending on the relation between the $B$-field mass and the radius of the extra dimension. In string theory, it might not be possible to engineer exponentially sequestered sectors (e.g.~\cite{Kachru:2007xp,Berg:2010ha,Heckman:2019bzm}) as in phenomenological RS setups and as such these light modes may always be present in quantum gravity constructions.
    \item {\bf Large lattice index:}
    It has been shown that the lattice WGC does not hold in full generality but that a sublattice version holds in all known examples. If the index of the sublattice could be made arbitrarily large then the kinetic mixing signal estimated here could be suppressed. However, it is believed that there is a universal upper bound on the index and indeed in all known examples, it is an $\mathcal{O}(1)$ number.

\end{itemize}

\section{Conclusion and Outlook}
\label{sec:conclusions}

String theory provides a natural framework for exploring generic expectations at low energies since it provides us with a large number of consistent vacua. It is important to quantify these expectations by studying Landscape constructions, however these constructions tend to be limited to highly symmetric scenarios and can sometimes lead to biased results. A complementary approach is to appeal to a feature believed to hold in quantum gravity generally and directly derive estimates from it. In this work, we performed such an estimate by considering the connection between the WGC and kinetic mixing. Ideally, both approaches should give comparable results and we verify this by also computing the kinetic mixing in a large number of heterotic orbifolds and Type IIB supergravity on a Calabi-Yau manifold.

In this work, we focused on the massless dark photon, but our results are equally applicable to the case of massive light $U(1)$ gauge bosons as well. We generically expect the WGC to apply to light gauge bosons. In particular, since these bosons get masses from either Higgsing or the Stueckelberg mechanism, there is some UV scale where they are effectively massless. Above this scale, the theory contains a massless dark photon and the analysis we carried out holds. To connect these results to experiment, we require knowledge of the dynamics governing the mass generation mechanism, which are model-dependent, but calculable.

A clear future direction that would allow for further explorations of this type is to better understand the distribution of massive states. On the one hand, some of these states could reside in the dark sector of our universe. Alternatively, they could contribute threshold effects to the low energy theory when integrated out. For instance, in the example of kinetic mixing we consider here, we saw that there is a minimum variance to kinetic mixing as a function of the gauge coupling. It would be interesting to see if this effect is verified in more rigorous constructions or seen using mass distributions derived from string theory. Genericity studies of this type could also help in isolating a few `loophole' scenarios for focused exploration if experiment happens to rule out all generic expectations.

\acknowledgments{We thank Prateek Agrawal, Cora Dvorkin, Matthew Reece, Cumrun Vafa and Weishuang Linda Xu for useful discussions and comments. We are also grateful to Keith Dienes and Sa\'{u}l Ramos-S\'{a}nchez for email correspondence. The work of GO is supported in part by a grant from the Simons
Foundation (602883, CV). AP is supported in part by an NSF Graduate Research Fellowship Grant DGE1745303, the DOE Grant DE-SC0013607, and the Alfred P. Sloan Foundation Grant No.~G-2019-12504.
For some of our results on heterotic orbifolds, we found the \href{https://orbifolder.hepforge.org/}{Orbifolder Tool} useful.}

\appendix
\section{Analytic Estimate of Mixing}
\label{sec:appendix_ratio_dist}

The calculation of $\langle \chi^{2} \rangle$ can be broken into a few steps. In order to compare with the string constructions, we chose a particle spectrum that included states with charges $( Q_a, \pm |Q_b|)$. This led to the contribution we see in Equation~\ref{eqn:paircontribution}. The first step then is to find the distribution of $r = c_{+}/c_{-}$. In statistics, this is known as the ratio distribution. For two random variables $X$ and $Y$, the distribution of $R = X/Y$ is given by
\begin{equation}
P_{R}(R) = \int_{0}^{\infty}f_{y}(y)f_{x}(Ry)ydy
\label{eqn:ratio_dist}
\end{equation}
where we have made use of the fact that $c_{+}$ and $c_{-}$ are independent variables so their joint PDF is just the product of their individual PDFs. To compute this ratio distribution, we have to consider the cases $r > 1$ and $r < 1$ separately. For $r < 1$, we integrate over a triangle in the $c_{+} - c_{-}$ plane defined by $c_{+} \in [0, rc_{-}]$ and $c_{-} \in [0, 1]$. Computing the integral in Equation~\ref{eqn:ratio_dist}, we find
\begin{equation}
    P_{r}(r) = \frac{1 + e^{(1+r)\beta}(-1 + \beta + r\beta)}{(1+r)^{2}(e^{\beta}-1)^{2}} \qquad r < 1
\end{equation}
The $r > 1$ case is slightly more involved. The region we integrate over is a trapezoid. We can break this into two regions. The first is defined by $c_{+} \in [0, rc_{-}]$ and $c_{-} \in [0, r^{-1}]$. The second is defined by $c_{+} \in [0, 1]$ and $c_{-} \in [r^{-1}, 1]$. Finding the CDF and differentiating with respect to $r$, we find
\begin{equation}
    P_{r}(r) = \frac{r + e^{\beta(1+r^{-1})}(r(\beta-1)+\beta)}{(1+r)^{2}r(e^{\beta}-1)^{2}} \qquad r > 1
\end{equation}
Given the probability distribution of $r$, we can find the probability distribution of $\chi$ which is a function of $r$. Suppose we take
\begin{equation}
    \chi = -\frac{1}{k}\log(r) \quad \to \quad r = \exp(-k\chi) \qquad k = \frac{48\pi^{2}}{g_{a}g_{b}Q_{a}Q_{b}}
\end{equation}
If we take $\chi = g(r)$, then we have
\begin{equation}
    P_{\chi}(\chi) = P_{r}\left(g^{-1}\left(\chi\right)\right)\left|\frac{d}{d\chi}g^{-1}(\chi)\right| = \left|k\right|\exp\left(-k\chi\right)P_{r}\left(\exp\left(-k\chi\right)\right)
\end{equation}
Symmetrizing with respect to $r$ and $r^{-1}$, we find the results in Equation~\ref{eqn:chi_dist}. With $P_{\chi}(\chi)$ in hand, we can compute $\langle \chi_{ij}^{2} \rangle$ by evaluating
\begin{equation}
    \langle \chi_{ij}^{2} \rangle = \int_{-\infty}^{\infty} \chi_{ij}^{2}P(\chi_{ij})d\chi_{ij} = \int_{-\infty}^{0} \chi_{ij}^{2}P(\chi_{ij})d\chi_{ij} + \int_{0}^{\infty} \chi_{ij}^{2}P(\chi_{ij})d\chi_{ij}
\end{equation}
$P_{\chi}(\chi)$ is piecewise defined, so we split our integral into two at $\chi = 0$, which corresponds to $r = 1$. Since $P_{\chi}(\chi) = P_{\chi}(-\chi)$ and the integrand is even, we can consider just one of these regions, which provides us the computational benefit of having to only consider $r > 1$ \emph{or} $r < 1$. Computing this integral gives us the results in Equation~\ref{eqn:chiij_variance}. Finally, to compute $\langle \chi^{2} \rangle$, we have to integrate $\langle \chi_{ij}^{2} \rangle$ over the $\beta < 1$ ellipse. Since we have already taken into account states with charges $\pm |Q_b|$, we focus on the half-ellipse where $Q_b > 0$. Integrating over this region, we arrive at the results in Equation~\ref{eqn:chi_variance}.

\section{CY 3-form bases and Siegel's Fundamental Domain}
\label{app:siegel}
We begin by recalling the symplectic basis transformations that act on the real 3-form basis of a CY manifold. For a CY manifold with Hodge number $h^{2,1}$, there are $2(h^{2,1}+1)$ 3-forms in real cohomology that we label $(\alpha_i, \beta^j)$ with $i,j = 0, \ldots, h^{2,1}$. These have the following pairing relations:
\begin{align*}
    \int_{\rm CY} \alpha_i \wedge \beta^j = \delta_i^j; \qquad
    \int_{\rm CY} \alpha_i \wedge \alpha_j = 0; \qquad
    \int_{\rm CY} \beta^i \wedge \beta^j = 0
\end{align*}
which are preserved by transformations under the symplectic group $Sp(2h^{2,1} + 2; \mathbb{Z})$. For a matrix $\begin{pmatrix}
A & B \\ C & D
\end{pmatrix}\in Sp(2h^{2,1} + 2; \mathbb{Z}) $, the action on the $(\alpha_i, \beta^j)$ and the complex matrix $\mathcal{M}$ defined in~\ref{eq:Mmatrix} is given by:
\begin{align*}
    \begin{pmatrix}
        \alpha \\ \beta
    \end{pmatrix} \rightarrow
    \begin{pmatrix}
        A & B \\ C & D
    \end{pmatrix} \cdot
    \begin{pmatrix}
        \alpha \\ \beta
    \end{pmatrix}; \qquad
    \mathcal{M} \rightarrow (A \mathcal{M} + B)(C \mathcal{M} + D)^{-1}.
\end{align*}
As briefly mentioned previously, we work with the matrix $\mathcal{M}^*$ which has the same transformation as above with the replacement $\mathcal{M} \rightarrow \mathcal{M}^*$. The matrix $\mathcal{M}^*$ takes values in Siegel's upper half space of genus (or degree) $h^{2,1} + 1$ defined to be the subset of $(h^{2,1} + 1) \times (h^{2,1} + 1)$ complex matrices that are symmetric with a positive definite imaginary part. The latter condition ensures that ${\rm Im}\mathcal{M}$ is negative definite which is required for the gauge kinetic terms to have the correct sign. The symplectic action preserves these conditions. As a familiar example, one can consider the genus 1 case of Siegel's upper half space which is the well-known Teichm\"{u}ller space describing the complex structure of the torus with the usual $PSL(2,\mathbb{Z})$ action.

As in the case of the upper half plane, a natural question to ask is: what is the fundamental domain of Siegel's half space under the symplectic action? It turns out that the conditions defining one such fundamental domain provide a choice of electric-magnetic duality frame that matches our needs. This also resolves the issue related to basis-dependence. We now specialize to the case of interest, i.e. genus 2, and summarize these conditions (see for example~\cite{Siegel,dupont2006moyenne, jaber2017algorithmic} for a more detailed account) and their physical meaning.\\

\noindent {\bf Definition (Siegel's fundamental domain for genus 2): } Let $\mathcal{M}_{ij} = X_{ij} + i Y_{ij}$ be a $2 \times 2$ complex matrix that is symmetric, with $Y$ positive definite. Then $\mathcal{M}$ is said to be in the fundamental domain if:
\begin{enumerate}
    \item $|X_{ij}| \leq \frac12$
    \item The matrix $Y$ has elements that satisfy $Y_{11} \geq Y_{00} \geq 2 Y_{10} > 0$
    \item $|{\rm det}(C \Omega + D)|\geq 1$ for matrices $C$ and $D$ that are submatrices of a symplectic matrix, as shown above.
\end{enumerate}

Given that the real part of $\mathcal{M}$ is the coefficient of the topological terms in the 4D action, the first condition amounts to using the periodicity of the $\theta$ angles so that they are not arbitrarily large. The second condition guarantees that the first $U(1)$ has a smaller gauge coupling than the second and that the mixing is small compared to the gauge couplings. Finally, the third condition ensures that the matrix $\mathcal{M}$ gives the most weakly coupled theory, although this is harder to demonstrate without considering the details of the algorithm provided in~\cite{dupont2006moyenne}.

The algorithm to reduce a matrix to Siegel's fundamental domain proceeds by repeatedly carrying out a few simple operations. The first operation is Minkowski reduction which is simply a basis change not related to electric-magnetic duality. Using integer coefficients, the fields are redefined to ensure that the two $U(1)$'s are as orthogonal as possible with gauge couplings and kinetic mixing satisfying the second condition in the definition above. This is followed by shifting the theta angles to ensure that the real part of the matrix lies in the range $[-\frac12, \frac12]$. These latter transformations are the analogue of $T$ transformations for the genus 1 case. Finally, one has to ensure that the third condition is satisfied. Na\"{i}vely, this seems to require checking an infinite number of inequalities, however Gottschling~\cite{GOTTSCHLING1959} showed that one only needs to verify this for a finite set of 19 matrices. Correspondingly, there is a series of 19 transformations (analogous to the $S$ transformations of genus 1) that are carried out whenever one of Gottschling's conditions is not satisfied. This algorithm is iterated until all three conditions are simultaneously satisfied.

\bibliographystyle{JHEP.bst}
\bibliography{ref.bib}
\end{document}